# One-dimensional disordered photonic structures with two or more materials

Alessandro Chiasera[a], Luigino Criante[b], Stefano Varas[a], Giuseppe Della Valle[c,d], Roberta Ramponi[c,d], Maurizio Ferrari[a,e], Lidia Zur[a,e], Anna Lukowiak[f], Ilka Kriegel[g], Michele Bellingeri[h], Davide Cassi[h], Francesco Scotognella[b,c,]*

[a]IFN-CNR CSMFO Lab, and FBK Photonics Unit, Trento, Italy; [b]Center for Nano Science and Technology@PoliMi, Istituto Italiano di Tecnologia, via Giovanni Pascoli, 70/3, 20133, Milano, Italy; [c]Politecnico di Milano, Dipartimento di Fisica, Piazza Leonardo da Vinci 32, 20133 Milano, Italy; [d]Istituto di Fotonica e Nanotecnologie CNR, Piazza Leonardo da Vinci 32, 20133 Milano, Italy; [e]Centro Studi e Ricerche Enrico Fermi, Rome, Italy; [f]Institute of Low Temperature and Structure Research PAS, Wroclaw, Poland; [g]Department of Nanochemistry, Istituto Italiano di Tecnologia (IIT), via Morego 30, 16163 Genova, Italy; [h]Dipartimento di Scienze Matematiche, Fisiche e Informatiche, Università di Parma, Viale G.P. Usberti n.7/A, 43100 Parma, Italy
*francesco.scotognella@polimi.it; phone 39 02 2399-6056; fax 39 02 2399-6126; www.fisi.polimi.it/en/people/scotognella

**ABSTRACT**

Here we would like to discuss the light transmission modulation by periodic and disordered one dimensional (1D) photonic structures. In particular, we will present some theoretical and experimental findings highlighting the peculiar optical properties of: i) 1D periodic and disordered photonic structures made with two or more materials [1,2]; ii) 1D photonic structures in which the homogeneity[3] or the aggregation[4] of the high refractive index layers is controlled. We will focus also on the fabrication aspects of these structures.

**Keywords:** one-dimensional photonic structures, disordered photonics, structure-property relationship.

## 1. INTRODUCTION

Photonic crystals are structures in which the periodic arrangement of materials with different refractive indexes results in a peculiar behaviour, characterized by the occurrence of the photonic band gap[5–7]. The photonic band gap is an energy region in which light is not allowed to propagate through the photonic structure. One-dimensional (1D) photonic crystals are a monodimensional alternation of the high and low refractive index materials and for these structures the photonic band gap occurs only in the direction of such alternation. 1D structures can be useful for several applications, as for example sensing[8–11], microcavity-based or distributed feedback lasers[12–14], solar cells[15–17]. The structure-property relationship in one-dimensional (1D) photonic structures, especially when they are characterized by aperiodic (Fibonacci, Thue-Morse, Rudin-Shapiro, Cantor, etc.) or disordered sequences, is an attractive field in the community of optics and photonics[18–22]. The possibility to fabricate 1D photonic structures with diverse and flexible techniques, as for example spin coating[16,23,24], radiofrequency sputtering[25,26] and pulsed laser deposition[27], allows to study these structures with a desired arrangement.

In this manuscript, we discuss the optical properties, in terms of light transmission, of 1D multilayer photonic structures in the cases of periodicity and disorder. We will mainly focus on the particular optical properties, from a theoretical and experimental point of view of: i) 1D periodic photonic crystals and 1D disordered photonic structures made with materials and made with more than two materials; ii) 1D disordered photonic structures in which the homogeneity, of the high refractive index layers distributed along the length of the structure, or the aggregation of the high refractive index layers among the low refractive index layers is controlled by certain models.

## 2. METHODS

In order to simulate the light transmission of the 1D photonic structures we employ the transfer matrix method, a versatile method already described in [21,28–30]. In the simulation the light impinges orthogonally the glass substrate (with refractive index $n_s$=1.46), the multilayer and air (with refractive index $n_0$=1). The parameters of the different multilayer structures studied here are indicated in the Results and Discussion section.

## 3. RESULTS AND DISCUSSION

In Figure 1 we show the sketches of 1D periodic, aperiodic and disordered photonic structures. The aperiodic 1D structure follows the Fibonacci sequence ABAABABAABAABABAABABA.

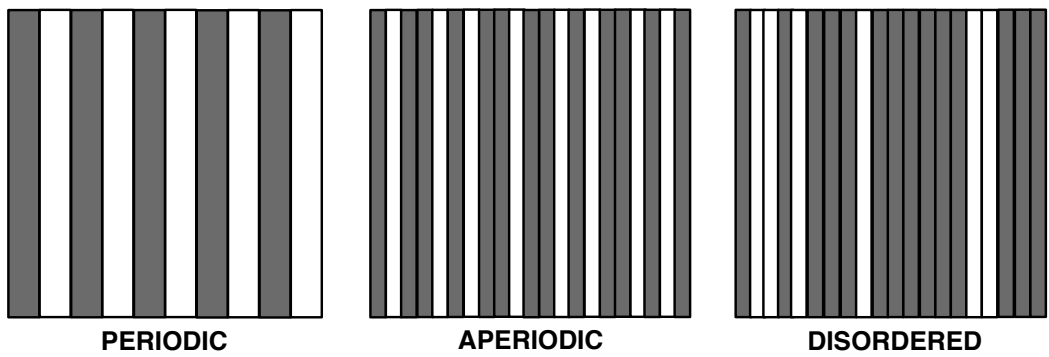

Figure 1. Sketches of 1D (a) periodic, (b) aperiodic (following the Fibonacci sequence ABAABABAABAABABAABABA) and (c) disordered photonic structures.

To make a photonic crystal with more than two materials we have to build structures with unit cells that include all the materials. For example, a three-material photonic crystal of $n$ unit cells follows a sequence $(ABC)_n$, a four-material photonic crystal of $n$ unit cells follows a sequence $(ABCD)_n$, etc.

We take into account the wavelength dispersions of the refractive indexes $n(\lambda)$ of different materials to make a reliable simulation of the 1D photonic structures. In Figure 2a we show the wavelength dispersions of the refractive indexes of $SiO_2$ [31], $Al_2O_3$ [32], $Y_2O_3$ [33] and $ZrO_2$ [34]. By employing these refractive indexes in the transfer matrix method we have simulated the transmission spectra of a 60 layer two-material ($SiO_2$ and $Y_2O_3$) photonic crystal (30 unit cells with 2 materials), of a 60 layer three-material ($SiO_2$, $Al_2O_3$ and $Y_2O_3$) photonic crystal (20 unit cells with 3 materials), and of a 60 layer four-material ($SiO_2$, $Al_2O_3$, $Y_2O_3$ and $ZrO_2$) photonic crystal (15 unit cells with 4 materials). The transmission spectra, in the energy range 0.4 – 2.2 eV, are depicted in Figure 2b. The thickness of each layer is a numerical value divided by the refractive index of the material at about 2 eV. Thus, the thickness of the $SiO_2$ layers is (240/1.457) nm, the thickness of the $Al_2O_3$ layers is (240/1.7665) nm, the thickness of the $Y_2O_3$ layers is (240/1.9264) nm, and the thickness of the $ZrO_2$ layers is (240/2.1535) nm. It is evident that, for structures made with $m$ materials, we observe $m$-1 photonic band gaps occurring. We would like to mention that an experimental fabrication of a ternary photonic crystal made, by solution processing, with cellulose, poly vinylalcohol and poly (N-vinylcarbazole) has been performed by Manfredi et al. [35].

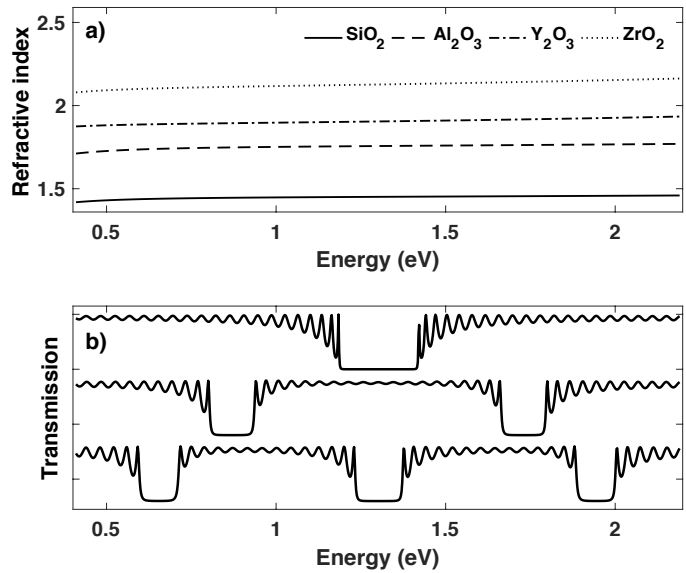

Figure 2. (a) Wavelength dispersions of the refractive indexes of $SiO_2$ (solid curve), $Al_2O_3$ (dashed curve), $Y_2O_3$ (dot-dashed curve), $ZrO_2$ (dotted curve). (b) Transmission spectra of (above) a photonic crystal made with $SiO_2$ and $Y_2O_3$, (center) a photonic crystal made with $SiO_2$, $Al_2O_3$ and $Y_2O_3$, (below) a photonic crystal made with $SiO_2$, $Al_2O_3$, $Y_2O_3$ and $ZrO_2$.

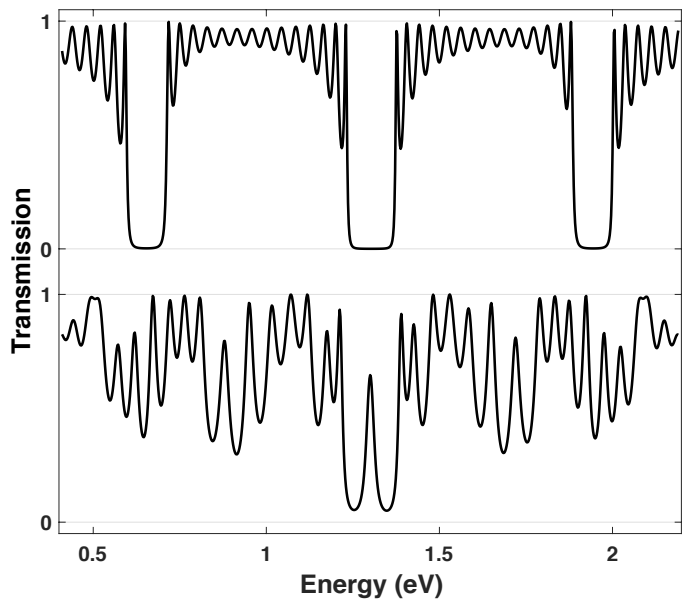

Figure 3. (above) Photonic crystal made with $SiO_2$, $Al_2O_3$, $Y_2O_3$ and $ZrO_2$; (below) disordered photonic structure made with $SiO_2$, $Al_2O_3$, $Y_2O_3$ and $ZrO_2$.

In Figure 3 we show the transmission spectrum of a four-material photonic crystals (above) and the spectrum of a four-material photonic structure where the 4 materials are randomly arranged along the structure.

In disordered photonic structures it is interesting to correlate the homogeneity of the structure with the optical properties. To measure the homogeneity of the structure we employ the Shannon index, that measure how many different species (in our case the high refractive index layers) are in a dataset and simultaneously takes into account the distribution of the species [3,21,36]. As a quantity to study the properties of photonic structures as a function of the homogeneity, we use the normalized total transmission, where the total transmission is the numerical integral of the light transmission in a selected range of wavelengths, and the normalized total transmission is the total transmission normalized by a certain value (in this case the total transmission of a periodic photonic crystal).

In Figure 4 we show the normalized total transmission (in the range 660 nm – 1760 nm) as a function of the Shannon index for a 64 layer 1D photonic structure made of SiC (as high refractive index material) and $SiO_2$ (as low refractive index material), 16 layers of SiC and 48 layers of $SiO_2$. Each layer has a thickness of 69 nm. We calculate the Shannon index, and simulate the corresponding total transmission with the transfer matrix theory, of 2500 disordered structure and the periodic crystal (+ sign in Figure 5).

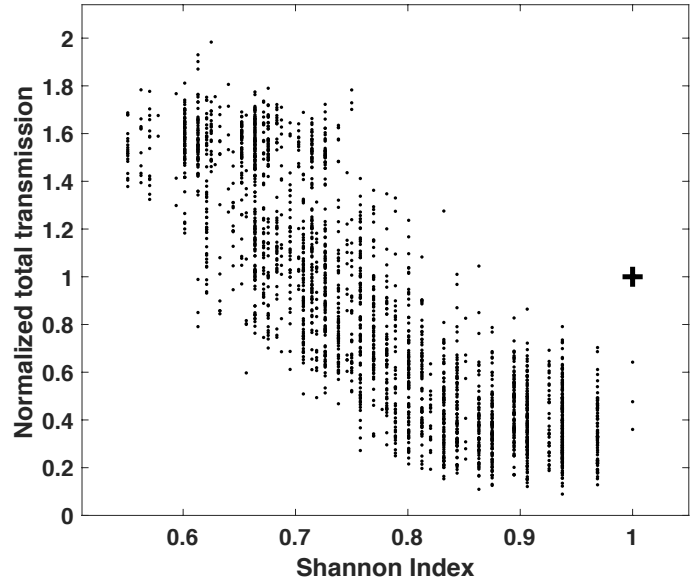

Figure 4. Normalized average transmission (normalized by the total transmission of the periodic photonic crystal) of $SiO_2$/SiC 1D photonic structures with different homogeneity. The + sign represents the transmission of the periodic photonic crystal.

The Sellmeier equation related to the wavelength dispersion of the refractive index of SiC is, in the range 0.425 – 2 μm (the wavelength $\lambda$ is in μm):

$$n^2_{SiC}(\lambda) = 0.21 + \frac{10.31\lambda^2}{\lambda^2 - 0.02635} + \frac{0.2231\lambda^2}{\lambda^2 - 0.02722} + \frac{105.7\lambda^2}{\lambda^2 - 410.4} \tag{1}$$

(fit from the experimental data in Refs. [37,38], with $R^2$ = 0.968). The Sellmeier equation for $SiO_2$ has been taken from Refs. [31,37].

It is interesting that the homogeneity of the 1D structure strongly affects the normalized total transmission. Disordered 1D photonic structures where the high refractive index layers are homogeneously distributed in the structures have a total transmission that is lower with respect to the period crystal. By decreasing the homogeneity of the structure we observe an increase of the total transmission: e.g. structures with the Shannon index <0.6 have a total transmission that is higher with respect to the one of the periodic crystal.

Finally, we can also correlate the the aggregation of the high refractive index layers in the 1D structure with the optical properties of such structures. We have observed that, for distributions of clusters that follow a power law, if we plot the total transmission as a function of the exponent of the power law, such trend follows a sigmoidal function [4].

## 4. CONCLUSION

In this manuscript we have discussed the light transmission of different types of 1D photonic structures. We have studied 1D periodic and disordered photonic structures made with more than 2 materials. Then, we have correlated the optical properties of 1D disordered photonic structures to the homogeneity of the structures, in terms of distribution of the high refractive index layer between the low refractive index layers, and to the aggregation of high refractive index layers in the structure. These findings can be interesting, from a fundamental point of view, for a better understanding of the the properties of disordered structures, and, from the applied point of view, for the fabrication of optical filters.

## ACKNOWLEDGEMENT

This project has received funding from the European Union's Horizon 2020 research and innovation programme (MOPTOPus) under the Marie Skłodowska-Curie grant agreement No. [705444], as well as (SONAR) grant agreement no. [734690]. The research activity is partially performed in the framework of COST Action MP1401 Advanced fibre laser and coherent source as tools for society, manufacturing and lifescience (2014-2018), ERANET-LAC FP7 Project RECOLA - Recovery of Lanthanides and other Metals from WEEE (2017-2019) and Centro Fermi MiFo "Microcavità Fotoniche" (2017-2020) project.